# Neutrinos and the challenges of particle physics

**José W. F. Valle**
*AHEP Group, Institut de Física Corpuscular --*
*CSIC/Universitat de València, Parc Científic de Paterna.*
*C/ Catedrático José Beltrán, 2 E-46980 Paterna (Valencia) - SPAIN*
*E-mail: valle@ific.uv.es*

I present a grand view of some of the main particle physics challenges from a neutrino perspective. After a brief review of the current status of neutrino physics, I illustrate the possible role neutrinos can play in the elucidation of various puzzles in particle physics and cosmology, for which the Standard Model offers no answer.

# 1. Introduction

The long-awaited discovery of the Higgs boson was exciting enough to bring the field of particle physics into euphoria. Despite its great significance, however, it does not represent the last brick in the construction of particle physics. The historic discovery of neutrino oscillations is as important, and likely to herald the beginning of a new chapter in the history of our field. Indeed, it may shed light upon several open issues in particle physics associated with the mass generation problem, the family problem, the unification of fundamental forces of nature and the way quarks and leptons get organized. In addition, there are several cosmological puzzles for which neutrino physics may provide a key input. These are associated, for example, with the understanding of the nature of dark matter and the origin of the baryon asymmetry of the universe. In short, neutrinos may have potentially important implications to our microscopic understanding of the universe. Here I illustrate these points with concrete examples, showing how neutrino physics may meet some of the challenges facing particle physics and cosmology.

## 1.1 Status of neutrino oscillations 2018

The bulk of experimental neutrino data is well described by the oscillation hypothesis. In addition to the solar and atmospheric neutrino data, the reactor and long baseline accelerator samples provide not only independent confirmation of the oscillation phenomenon, but also an improved parameter determination. We have also had atmospheric neutrino studies from high energy neutrino telescopes such as ANTARES and IceCube DeepCore. A fully consistent global picture emerges, the three-neutrino paradigm, illustrated in Fig.1 (see [1] for details). This figure summarizes the status of neutrino oscillations just prior to the Neutrino conference this summer.

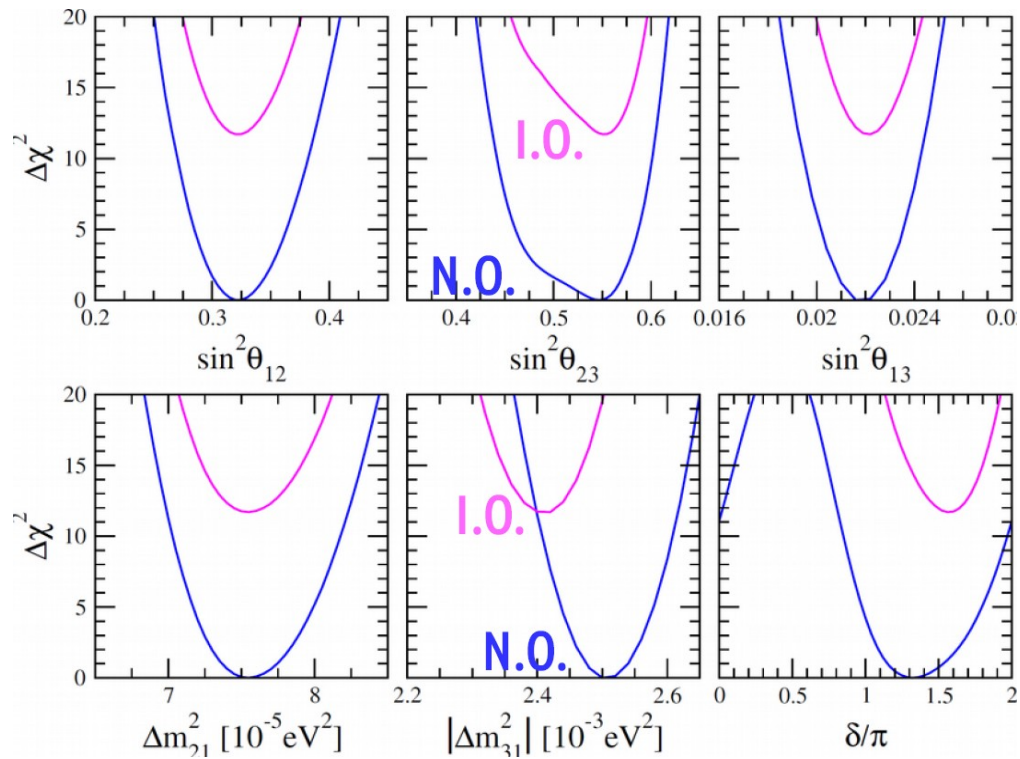

Fig.1 Status of neutrino oscillation parameters, 2018, from Ref.[1]

One sees that solar and atmospheric angles are found to be surprisingly large in comparison with their quark sector counterparts. In fact, the smallest of the lepton mixing angles, the reactor angle, is similar in magnitude to the largest of the CKM angles, namely the Cabibbo angle. Here we may have a hidden message of nature! Some have speculated that, perhaps, this is not a mere coincidence, and that there may be a common origin for quark and lepton mixing parameters [2]. Concerning mass splittings, they look tiny on the scale of charged fermion masses in the standard model, reinforcing that neutrinos are special. A new result is the 3σ preference for normal mass ordering, indicated by the lower (blue) contours. The results corresponding to inverted mass ordering are shown in magenta. Finally, one also has an improved hint of leptonic CP violation, the sensitivity coming mainly from the T2K data sample. The global oscillation analysis is in good agreement with those of Bari and NuFit.

In summary, two of the oscillation parameters remain poorly measured, namely $\theta_{23}$ and $\delta$. There are good long-term prospects for improving the measurement of these parameters in future experiments, such as DUNE [3-6], except for the octant discrimination, which will remain poor if the true value happens to be close to maximality [7]. By and large, however, neutrino physics has reached the precision era. It is nevertheless mandatory to probe robustness of the oscillation hypothesis and this requires to go beyond the three-neutrino paradigm. This is also of great theoretical importance, as one expects most theories of neutrino mass generation to bring in sub-leading effects, such as unitarity violation and non-standard interactions [8-10]. These can substantially downgrade [11,12] the degree with which future experiments will explore the $\theta_{23}$-$\delta$ plane. However, they also open a new window of opportunity to delve into the origin of neutrino mass by probing the associated mass scale through the study of these deviations [13,14].

### 1.2 Neutrinoless double-beta decay

Oscillation studies do not probe absolute neutrino masses. The first relevant tool here is to study beta decay endpoint spectra, as in the Katrin setup (Drexlin's talk). Neutrinoless double-beta decay searches provide a complementary probe. Moreover, they also serve to test the Majorana nature of neutrinos. In addition to the three neutrino masses, the amplitude involves the mixing parameters measured in neutrino oscillations and the two Majorana phases [8][1], neatly expressed in the symmetrical parametrization of the lepton mixing matrix as [15],

$$|c_{12}^2 c_{13}^2 m_1 + s_{12}^2 c_{13}^2 m_2 e^{2i\phi_{12}} + s_{13}^2 m_3 e^{2i\phi_{13}}|$$

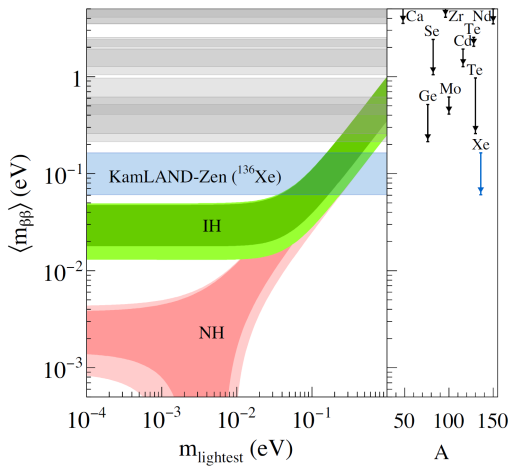

Fig 2: Expected neutrinoless double beta decay amplitude parameter versus experimental sensitivities, see Gando's talk.

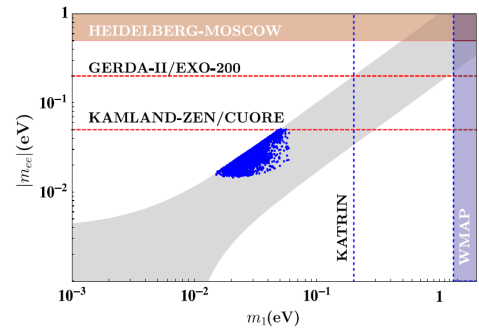

Fig 3: Flavour models typically lead to a lower bound on the neutrinoless double beta decay amplitude, even for the normal mass ordering case. This example is taken from Ref.[17]

---

[1] These can not affect lepton-number-conserving processes, such as conventional oscillations [16].

The two shaded branches labeled NH and IH in Fig.2 correspond to the cases of normal and inverted mass ordering, while the horizontal blue band illustrates the Kamland-zen sensitivity. Upcoming searches are designed to fully cover the inverted ordering region. Unfortunately, these experiments will not cover the normal region, currently preferred by oscillations, thanks to possible destructive interference that can arise in this case due to the Majorana phases. Here we note, however, that if the oscillation parameters are restricted beyond what we currently know from oscillation experiments, the cancellation can be prevented, leading to a lower bound even for the normal ordering case. This happens in a number of flavor-symmetry-based models [17], though the resulting bounds (Fig.3) are strongly model-dependent. One can, however, obtain a truly model-independent lower bound if the theory's neutrino content is "incomplete"[2], as seen in Fig.4. There we show the effective mass as a function of the Majorana phase. Normal and inverted orderings correspond to the light and dark green bands. Their width accounts for the uncertainties in the neutrino oscillation parameters. The solid black lines represent the current best fit values, while the horizontal bands collect the experimental limits and sensitivities.

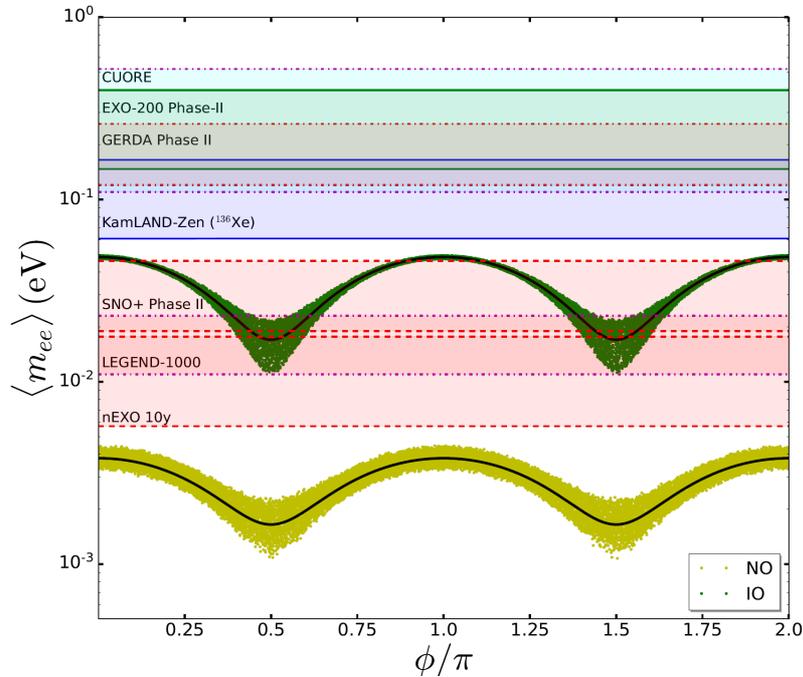

*Fig 4: Effective mass parameter as a function of the Majorana phase for normal (lower band) and inverted (upper band) mass orderings, in "incomplete" 321 scheme [18]. The widths of the shaded bands represent the current uncertainties in the neutrino oscillation parameters.*

To wrap up this discussion one can just say that, in addition to providing a powerful tool to probe small neutrino masses, neutrinoless double beta decay searches offer a model-independent test of the Majorana nature of neutrinos, through the black-box theorem [19].

## 2 Origin of small neutrino mass

Weinberg was first to note that one can add to the Standard Model a dimension-five operator involving two lepton doublets and two Higgs doublets. This turns into a Majorana neutrino mass after electroweak symmetry breaking. However, this is far from a complete

---

[2] Incomplete means less "right" than left-handed neutrinos, as can occur in SU(3)xSU(2)xU(1) seesaw [8]

theory of neutrino mass generation, to the extent that we have no clue as what is the underlying mechanism, its associated mass scale and flavour structure, or the coefficient sitting in front.

A consistent tree-level realization invokes the exchange of heavy "right" neutrinos at scale $v_1$ (type-I seesaw) or heavy scalar triplets (type-II seesaw) with a small induced vev $v_3$ [8] (note, the original seesaw nomenclature in [8] was opposite to today's). A *dynamical* explanation of the small neutrino mass in such *high-scale seesaw* arises if the masses of the "right" neutrinos come from a singlet vev $v_1$, while the heavy scalar triplet has vev $v_3$. Minimization of the Higgs potential is consistent with the vev hierarchy $v_3 \ll v_2 \ll v_1$, where $v_2$ is the Standard Model Higgs vev responsible for generating the W-boson mass. Such "123" seesaw also implies the existence of the majoron, associated to the spontaneous violation of lepton number [22]. Formulating the seesaw mechanism in terms of the SU(3)xSU(2)xU(1) gauge group, as opposed to its left-right extensions, defines the most general seesaw. Within this formulation we can have any number of "right" neutrinos, providing templates for new theories of neutrino mass, e.g. models with less "right" than left-handed neutrinos, such as the recent *scoto-seesaw* mechanism [20]. It also opens up the possibility of *low-scale seesaw* realizations [21], consistent with t'Hooft's naturalness principle. In such *low-scale seesaw* the neutrino mass mediators are Quasi-Dirac, and can lie within reach of high energy colliders. Non-trivial charged and neutral current mixing matrices lead to potentially large universality violation in neutrino propagation [9,10].

It is also worth noticing that the seesaw idea, both type-I as well as type-II, can be realized also if neutrinos are Dirac fermions, as a result of a protective symmetry. The associated dimension-5 and -6 operators have been classified [23], and there are a number of possible UV-completions with *Dirac seesaw mechanism,* in which the smallness of the Dirac neutrino mass is symmetry-protected, very much as in the Majorana case. Again, with low-scale realizations.

Finally, the smallness of neutrino mass may arise radiatively, through mediator particles that can lie at the few TeV scale, accessible to colliders. Typical *radiative mechanisms of neutrino mass generation* involve new scalars and fermions. An interesting example where gauge boson exchange generates neutrino masses radiatively is shown in Fig.5, from Ref.[24]. The gause boson running in the loop is required by the extended SU(3)xSU(3)xSU(1) electroweak gauge symmetry (331 for short). This theory was proposed long ago and provides a way to "explain" why the number of families is three, the same as the number of colors [25].

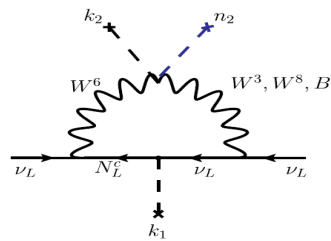

*Fig 5:Radiative Majorana neutrino mass from gauge boson exchange in 331 [24]*

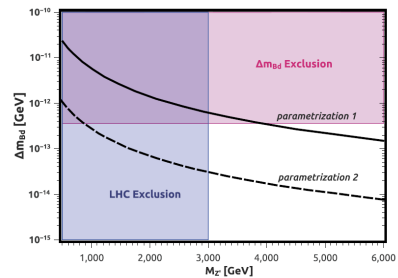

*Fig 6: Probing neutrino mass mediators in 331 theory [25].*

The mediator particles can lie at the few TeV scale, so they can lead to signature at colliders, for example, Drell-Yan production of new neutral gauge bosons in proton-proton collisions at the LHC, as well as new contributions to meson-anti-meson mass differences, as probed at Babar, Belle, LHCb, etc. The relevant sensitivities are illustrated in Fig.6, see [26] for details.

**3 Family symmetries and the flavor problem**

The Standard Model lacks an organizing principle to answer questions such as *Why is lepton mixing so special when compared with the CKM matrix characterizing the quark sector? Does the Cabbibo angle play any special role in the physics of flavor?* Indeed, understanding the family puzzle is one of the greatest challenges in particle physics.

The most powerful tool we have to organize the three quark and lepton families is symmetry. The simplest way to stack these families together is provided by the $A_4$ symmetry, that leads to maximal atmospheric mixing, zero reactor angle and an adjustable solar angle [27]. Back in 2002 this was a good prediction. Now it is no longer the case, after the results from reactor experiments Daya-Bay & RENO. Fortunately, the theory can be easily revamped [28] so as to generate a non-zero reactor angle and physical leptonic CP violation. This is achieved in a predictive way, correlating $\theta_{23}$ and $\delta$, as seen by the green band in Fig.7.

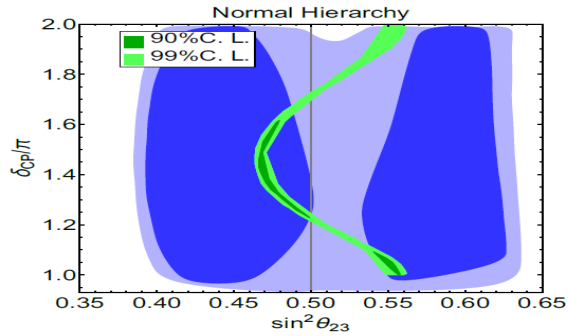

*Fig 7: The thin band is the predicted correlation in the model of Ref.[28]*

The figure shows the regions of three-neutrino oscillation parameters allowed at 90% and 99% C.L. from the unconstrained global fit (dark & light blue) and within the revamped $A_4$ scenario (dark & light green). The currently preferred case of normal mass ordering is assumed.

Correlations appear in many other models. For example, the curved band shown in Fig.8 is the theory prediction of Ref. [29], the vertical and horizontal bands result from neutrino oscillation global fits.

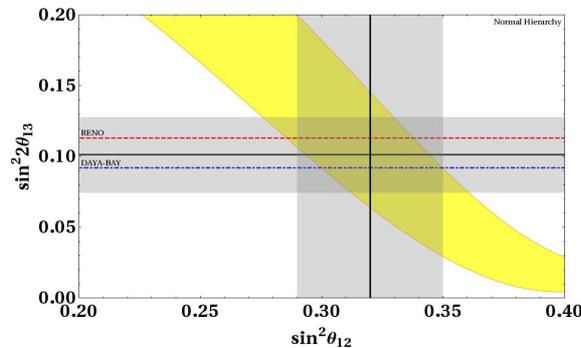

*Fig 8: Allowed oscillation parameters in the model of Ref. [29]*

A complete theory of flavor would require more than the ability to predict fermion mixing angles and CP violating parameters. The flavor challenge also includes the need to bring an organizing principle for the pattern of fermion masses themselves, which span many orders of magnitude, from the electron mass all the way to the top quark mass, not to speak of neutrinos. An interesting observation to be made here is that a number of flavor theories proposed to shed light on the neutrino oscillation parameters have nicely converged into a unique and successful *quark-lepton unification golden formula*

$$\frac{m_\tau}{\sqrt{m_e m_\mu}} \approx \frac{m_b}{\sqrt{m_s m_d}}.$$

One of these models is in Ref.[17]. Alternatives are given in [30-32]. This golden formula is reminiscent of bottom-tau unification in Grand unified theories. However, in contrast to that case, it only requires the Weinberg-Salam gauge group of the Standard Model, combined with a flavor symmetry. All families are involved and only mass ratios, so the relation is rather stable under renormalization. The good precise charged lepton and bottom quark mass measurements may be used to make predictions on the poorly known down- and strange-quark masses. It is remarkable that we reach such predictions starting only from our motivation to shed light on neutrino oscillation parameters! This illustrates the centrality of neutrinos in particle physics.

We now turn to a very interesting alternative theory scenario addressing both neutrino flavor predictions as well as mass hierarchies. The basic proposal is to combine the idea of *warping* due to Randal-Sundrum, with the imposition of *family symmetries*. This approach may be called *warped flavor model* and was proposed in [33]. In this case mass hierarchies are "predicted" by geometry, while mixing angle and CP phase predictions arise from the underlying family symmetry, in that case the $\Delta_{27}$ group, which leads to oscillation predictions.

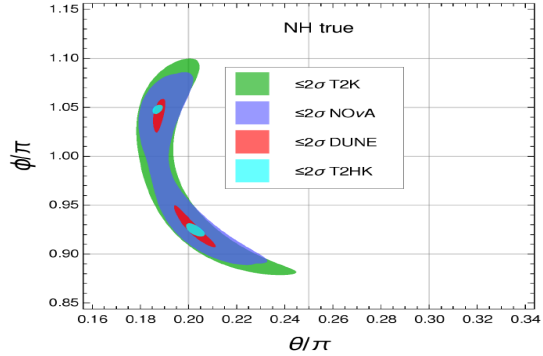

*Fig 9: Current and future oscillation constraints on warped SM, from [35]*

The 3 mixing angles and the CP phase are expressed analytically in terms of just two parameters $\theta$ and $\Phi$. By taking into account the predictions of the model [34] one can perform a *constrained* global fit of neutrino oscillation data, in order to determine the allowed regions of oscillation parameters [35]. Likewise, taking into account the design specifications of the upcoming long-baseline experiments, one can determine the future expected sensitivities. The results are shown in Fig.9. They show how the various sensitivity regions will substantially shrink after these experiments become available, especially DUNE and T2HK. Using the simple analytical formulae of the model, these results can be converted back into sensitivity regions in the physical oscillation parameters.

The above examples illustrate how to obtain neutrino parameters from first principles. The predictions emerge on a one-by-one basis, as a result of the assumed family symmetry at the basic Lagrangian level. A more model-indepedent symmetry-based approach can be envisaged in which all one cares are the symmetries of the mass matrices, irrespective of how they are generated from first principles. This approach exploits the invariance under flavor-generalized CP transformations and can lead to interesting predictions [36]. In particular, it can lead to a systematic way of revamping successful pattern such as mu-tau symmetry or TBM, so as to make them consistent with the latest oscillation global fits. An example is illustrated in Fig.10

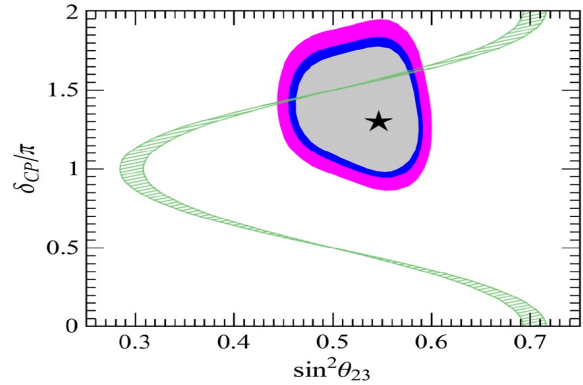

*Fig 10: Revamping the TBM ansatz through generalized CP symmetries, from [37]*

The correlation between the atmospheric angle $\theta_{23}$ and the CP phase $\delta_{CP}$ predicted by a generalized TBM matrix is given by the hatched band, while the 1, 2, and 3$\sigma$ regions allowed by the current generic neutrino oscillation global fit are indicated by the shaded areas.

**4 Neutrino pathways to unification**

The measured coupling constants associated to electromagnetic, weak and strong interactions show a trend towards converging, when evolved to high energies through the renormalization group equations, see Fig.11, left panel. This has been taken as suggestive of the idea of unification. Indeed, this works rather well in the context of unified supersymmetric theories. Unfortunately, however, there has been so far no evidence for super-partners at the LHC nor for proton decay signals at Super-Kamiokande.

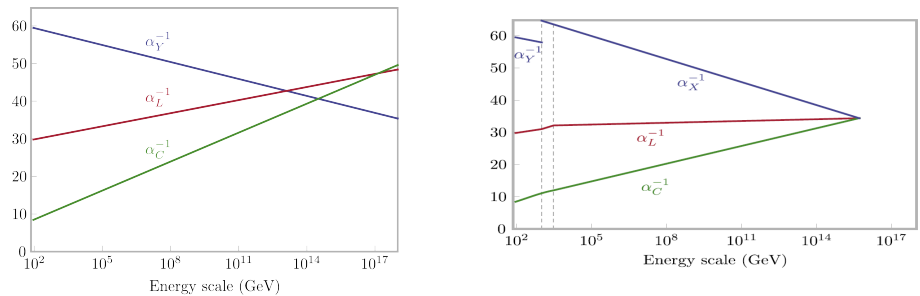

*Fig 11: Gauge coupling evolution in the Standard Model (left). The right panels shows how the same physics inducing neutrino mass may also cause gauge coupling unification [38]*

It could well hapen that the physics responsible for neutrino mass generation can also induce gauge coupling unification. As an example, a non-supersymmetric SU(3)xSU(3)xSU(1) theory was suggested in [38], containing three lepton octets, assumed to have 3 TeV mass in the right panel of Fig.11. The existence of such motivated "unification" setups seems suggestive, given the absence of signals associated with conventional supersymmetric unification.

## 5 Comprehensive unification

Standard unification scenarios employ gauge symmetries acting only within a given family, thus providing no link between different familes of quarks and leptons. As in the Standard Model these are just "replicated". Inspired by beauty of neutrinos in SO(10), in which each family (including the right-handed neutrino) sits neatly in the spinorial representaion, it has long been suggested that forces and families might be unified together within the framework of larger orthogonal groups. Spontaneous symmetry breaking would generate copies of quarks and leptons organized in family space. Because of the structure of orthogonal groups, however, the reduction of higher spinors necessarily produces unwanted mirror fermions.

A new approach has been recently proposed to rescue this program. It consists in promoting 4-dimensional Minkowski space into 5-dimensional Anti-de-Sitter space and using orbifold boundary conditions in order to decouple the mirror fermions, as illustrated in Fig.12 [39].

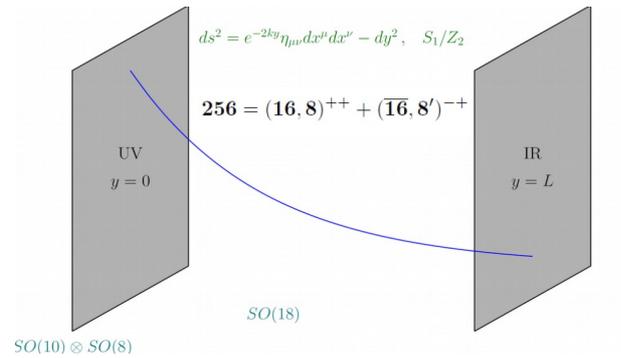

*Fig 12: Comprehensive unification approach [39]*

The setup starts from the 256 spinor representation in SO(18). One finds that, as a result of the orbifold boundary conditions, mirror fermions decouple after reduction, leaving eight chiral families. This number can be further reduced to three, since the new hypercolor interactions arising from the SO(5) subgroup of SO(8) may become confining at a few TeV.

How about flavor predictions? Within a bottom-up approach one may use the remaining SO(3) as family symmetry [40]. The construction is consistent only in the presence of a Peccei-Quinn symmetry, so the existence of the axion becomes essential in this setup, its scale being identified with the seesaw scale. Remarkably, one re-obtains the *quark-lepton unification golden formula* discussed in Sec.3, along with a dynamical framework for describing quark mixing and CP violation. Consistent CKM predictions are obtained, such as the Gatto-Sartori-Tonin relation for the Cabibbo angle.

## 6 Connection with cosmology

Precious information about the Big Bang comes from cosmic microwave background studies. These imply an upper bound on absolute neutrino masses, complementary to what is inferred from beta decay endpoint spectra as well as neutrinoless double beta decay searches (Wong's talk). From these studies one also infers the need for cosmological dark matter, whose detailed nature remains unknown. Even though neutrinos can make up at most 1% or less of this mysterious dark matter component, the physics associated to their mass generation may hold the key to the microscopic interpretation of the dark matter problem. Two ideas illustrating this fascinating connection are worth highlighting.

The first, proposed 25 years ago, is that the majoron is a dark matter particle (Heeck's talk), with a mass induced by gravitational effects [41]. Its decay arises from its tiny coupling to neutrinos, proportional to the neutrino mass, making it a naturally metastable dark matter particle. Constraints from the CMB and possibly from X- and gamma-ray signals associated to its late-decays have been investigated, as well as implications for structure formation [42].

The other idea, dubbed "scotogenic" approach, is that dark matter is a radiative neutrino mass messenger. It has many possible realizations [43]. For example, it has been recently suggested that dark matter can emerge today as a stable neutral hadronic thermal relics whose stability follows from the exact conservation of B-L symmetry, and hence from the Dirac nature of neutrinos. These get radiative masses from the exchange of colored dark matter constituents.

Many more ways of relating neutrinos to dark matter have been discussed. On the other hand, other cosmological problems such as the baryon asymmetry of the universe and inflation may also be closely connected to the neutrino mass generation physics [44].

Work funded by the Spanish grants SEV-2014-0398 and FPA2017-85216-P (AEI/FEDER, UE) and PROMETEO/2018/165 (Generalitat Valenciana). I am happy to thank all my collaborators involved in the research sketched here.